\documentclass[aps,prl,10pt,twocolumn,floatfix,superscriptaddress]{revtex4-2}
\usepackage{graphicx,amsmath,amssymb,bm,mathtools}
\usepackage[T1]{fontenc}
\usepackage[colorlinks=true,citecolor=blue,urlcolor=blue,linkcolor=blue]{hyperref}
\usepackage{microtype}
\newcommand{\ket}[1]{\lvert#1\rangle}
\newcommand{\bra}[1]{\langle#1\rvert}
\newcommand{\E}{\mathbb E}
\newcommand{\Id}{\mathbb I}
\newcommand{\HH}{\mathsf H}
\newcommand{\II}{\mathcal I}
\newcommand{\JJ}{\mathcal J}
\newcommand{\sE}{s_{\mathrm E}}
\newcommand{\dd}{\mathrm d}
\begin{document}
\title{A Warped Born Rule Turns an Observer into a Maxwell's Demon}
\author{Olumakinde Ogunnaike}
\affiliation{Department of Physics, The University of California, Berkeley, CA 94618, USA}
\date{\today}
\begin{abstract}
Quantum mechanics seems to be an incredibly delicate theory. Slight alterations to nearly any part of its structure can lead to huge qualitative differences and wildly unphysical predictions. Here, we explore the many-body consequences of replacing the Born rule $p(m)\propto\|P_m\ket\psi\|^2$ by a warped rule $p_b(m)\propto\|P_m\ket\psi\|^{2b}$. Specifically, in systems with conserved particle number, evolution punctuated by repeated local particle observations acquires a probabilistic bias over measurement histories with an effect controlled by the sign of $b-1$. For $b>1$, this bias generically drives the gas to collect into nearly frozen configurations. Similar to the action of Maxwell's demon, the simple act of observation seems to extract entropy from a closed system. For $b<1$, the bias is reversed, generically scouring out long-wavelength fluctuations to produce a hyperuniform gas.  We study this effect in a one-dimensional quantum gas, where these two phases are derived analytically for a particular measurement scheme and corroborated numerically for more general procedures with discrete-time projective-measurement circuit models using matrix product state simulations.  We additionally discuss how this bias manifests in a common technique that often employs this warped rule, the ``replica trick,'' implying qualitative differences between finite-replica and replica-limit calculations.
\end{abstract}
\maketitle

\begin{figure}[t]
\includegraphics[width=\columnwidth]{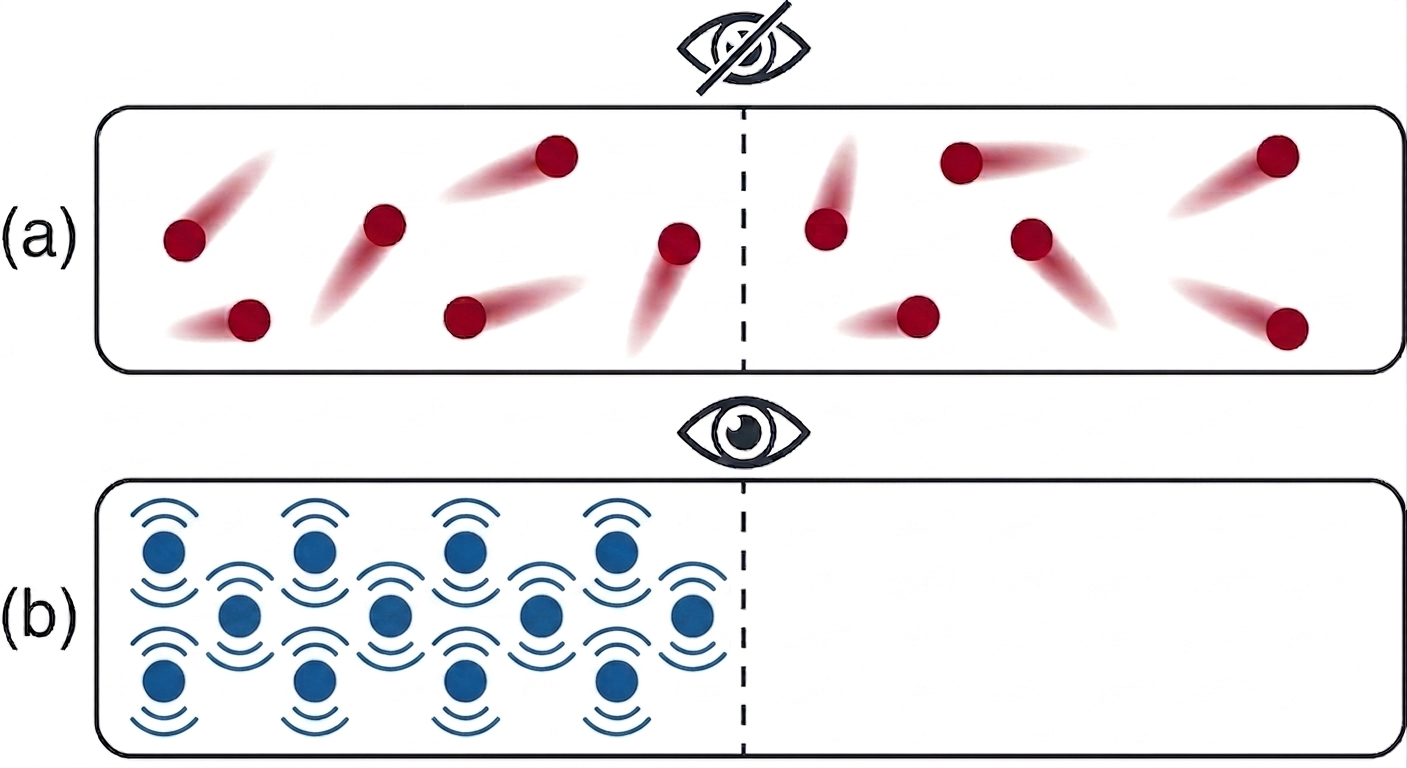}
\caption{A many-body Maxwell's demon. (a) Ordinary diffusion explores a uniform gas. (b) Reweighting occupation-measurement histories can collect the particles into dense, nearly frozen regions for warped rules with $b>1$.}
\label{fig:demon}
\end{figure}

Quantum mechanics may be an island in theory-space: deform one axiom, and the theory rapidly develops pathologies~\cite{AaronsonIsland}. Among the axioms, Born's rule occupies a special place: it is the bridge between the deterministic evolution of the state vector and the probabilistic outcomes of observation. A projective measurement with outcome $m$, described by a projector $P_m$, takes the state $\ket \psi$ to $P_m\ket\psi/\|P_m\ket\psi\|$, with probability $p(m) =\|P_m\ket\psi\|^2 $. A minimal deformation replaces this probability with
\begin{equation}
 p_b(m)=\frac{\|P_m\ket\psi\|^{2b}}{\sum_a\|P_a\ket\psi\|^{2b}},\qquad b>0,
 \label{eq:pborn}
\end{equation}
while leaving the normalized post-measurement state unchanged. On its face, it is a minimal change, touching only the weighting of outcomes and leaving Hilbert space alone. However, as many have explored before, the effects of this change are anything but minimal. Depending on the measurement postulates retained, known consequences include enhanced computational power, signalling, and failures of purification and local tomography~\cite{AaronsonPP,BaoBoulandJordan,GalleyMasanesClassification,GalleyMasanes}. 

Here, we explore overlooked collective effects of this warped rule in a many-body setting, focussing on the pedagogical example of a diffusing gas with conserved particle number. Left alone, its particles spread through the available volume. Born measurements preserve diffusion in the protocols below. For $b>1$, observations generically favor histories in which particles collect into dense regions, ``freezing'' (Fig.~\ref{fig:demon}). To see why, write the state's occupation-basis coefficients as $c_i$. Generic unitaries preserve $\sum_i|c_i|^2$, but not $\sum_i|c_i|^{2b}$ for $b\ne1$; unless we restrict them to mapping each definite occupation configuration to another, warped outcome weights require explicit normalization. Repeated observations turn this difference into a preference among histories.

Intuitively, this clustering is tied to the relative size of the warped norms. Born probabilities sum to one; warped weights sum to $N_b=\sum_m p_1(m)^b$, with signed normalization cost $\Phi_b=-\log N_b$, related to the entropy of that observation. For uncertain outcomes, $b>1$ gives $N_b<1$ and $\Phi_b>0$; $b<1$ reverses both inequalities. Raising Born probabilities to the power $b$ and renormalizing sharpens unequal probabilities for $b>1$, removing outcome entropy, and flattens them for $b<1$, adding it.

We can also see which alternatives this bias favors by comparing relative odds:
\begin{equation}
 \frac{p_b(a)/p_b(c)}{p_1(a)/p_1(c)}
 =\left[\frac{p_1(a)}{p_1(c)}\right]^{b-1}.
 \label{eq:intro_odds}
\end{equation}
Here $a$ and $c$ are alternative outcomes of the same observation, with the same preceding measurement record. When an outcome requiring a move is less likely than one leaving the particle in place, positive warping suppresses that move and negative warping enhances it. In the solvable model below, this becomes a bias in kinetic activity---the number of particle moves along a history. Reducing the entropy of successive measurements and suppressing motion provide the statistical origin of freezing at $b>1$.

These effects, and the bias behind them, have a familiar analogue. Maxwell's demon selectively lets fast particles through a door in one direction and slow particles in the other, establishing a temperature difference. This sorting reduces the gas's entropy and repeating the procedure requires the information retained and eventually erased by the demon~\cite{BennettDemon}. Much as the demon biases transport between sides of a box, the warped rule biases transitions between measurements. Repeated application selects some histories over others, with an analogous information-theoretic cost\footnote{This statistical cost is distinct from the  physical heat dissipated by a demon. No thermal reservoir or microscopic temperature is assigned to the gas.}. The observer has summoned a statistical Maxwell's demon simply by looking.

The same reweighting appears in normalized common-record replica calculations: one evolves $n$ copies of an unnormalized density matrix with the same measurement outcomes~\cite{BaoChoiAltmanTheory,LeonticaMcGinley,BaoChoiAltmanSymmetry,BentsenSahuSwingle,FengQuantumTrees,McGinleyHoMalz2025}. A measurement acts on these copies as
\begin{equation}
 \rho^{\otimes n}\mapsto\sum_m(P_m\rho P_m)^{\otimes n}.
 \label{eq:replica_update}
\end{equation}
For a fixed schedule, a record $\mathbf m$ with Born probability $P_{\rm B}(\mathbf m)$ contributes an n-copy trace $P_{\rm B}(\mathbf m)^n$. Normalizing the sum gives warped record weights with $b=n$: the statistics of independent copies conditioned on agreeing records. A single copy retains Born weighting. Calculations at fixed integer replica number have provided useful qualitative pictures in systems with discrete or no physical symmetries. With diffusing particles, however, this construction at $n>1$ can select qualitatively different dynamics from the properly Born-weighted $n \to 1$ replica limit, a concern emphasized by Barratt et al.~\cite{barratt_field_2022}. In a different setting, Feng et al. likewise find different Born and fixed-replica phase boundaries in a quantum-tree model~\cite{FengQuantumTrees}. Thus, this seemingly hypothetical deformation has concrete calculational relevance.

Below, we explicitly derive the bias behind these effects, relating it to the entropy of the measurement-record distribution, then show how one solvable monitored gas converts it into a kinetic activity bias driving a collective phase transition. Afterwards, we use matrix product state (MPS) simulations of discrete-time evolution with random measurements weighted by the warped rule to corroborate the existence of analogous phases in a more generic setting.

\textit{The local measurement basis.---}
A key ingredient in this bias is the local occupation basis. An observer measures local occupations $n_i=0,1$, defining particle configurations $C=(n_1,\ldots,n_L)$ and basis states $\ket C$. Starting from a resolved configuration $\ket C$, the Born transition probability is $|\bra{C'}U\ket C|^2$; the corresponding warped weight is $|\bra{C'}U\ket C|^{2b}$. Among two outcomes with unequal Born probabilities, this favors the more likely one for $b>1$ and the less likely one for $b<1$. Repetition compounds the preference, but the effect is clear after a single step. For a sufficiently small transition probability, $r$, positive warping favors staying put, while negative warping favors transitions. Schematically\footnote{Here $\mathcal{N}=(1-r)^b+2(r/2)^b$ and the bars illustrate the corresponding stay and transition weights schematically.}:

\begin{figure}[h]
    \includegraphics[width=0.95\columnwidth]{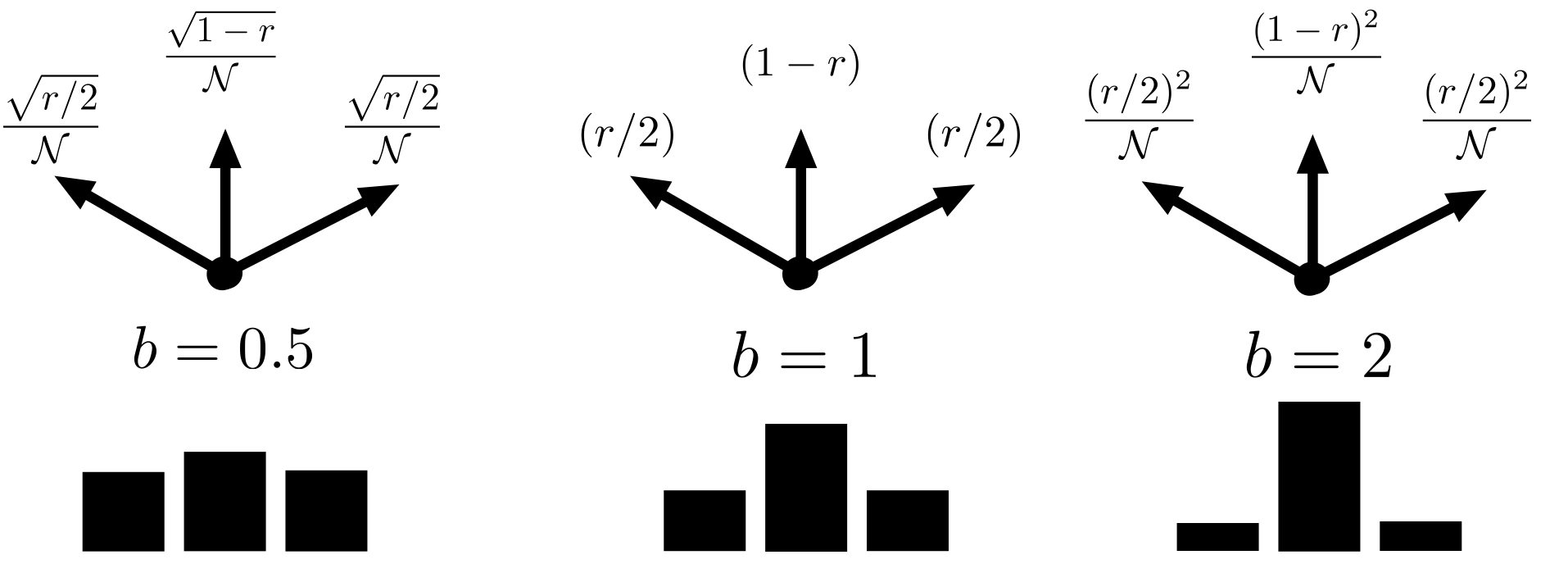}
    \label{fig:warped_step}
\end{figure}

Below, we specialize to the dynamics of a one-dimensional gas of conserved particles with intervening single-site occupation measurements. Slower motion alone need not produce clustering. In the solvable model below, packed configurations also offer fewer uncertain readouts, turning the record bias into a preference for dense domains. Measurements typically occur randomly at some fixed rate, so that the path between measurements is not of fixed length and a single readout generally leaves several configurations unresolved. However, for the solvable protocol discussed later, measurements occur after each unitary gate, and particle number conservation makes one site's outcome sufficient to resolve each update\footnote{In the case of sequential normalization discussed below, this exact resolution after each unitary provides a special case where normalization cancels the density-ordering bias.}. 

\textit{Monitored dynamics and entropic bias.---}
Consider a chain with single-site projectors $P_{i,1}=n_i$ and $P_{i,0}=1-n_i$. Between observations, unitaries conserving particle number produce long-wavelength diffusion with constant $D$. Measurements occur at each site at Poisson rate $\Gamma$.\footnote{Gate choices and measurement locations are externally prescribed, independently of outcomes. A fixed order is used for simultaneous readouts in discrete implementations.}

There are two reasonable ways to normalize the warped probabilities: globally after the full evolution, or sequentially at each measurement. Both sharpen or flatten Born probabilities in a similar fashion. 

Global normalization weights full trajectory records after evolution for a fixed time so that partial records depend on the summed weights of all measurements including future ones. Sequential normalization instead compounds local normalizations for each measurement, adding record-dependent terms to the bias. Let $\mathbf m=(m_1,\ldots,m_M)$ be the record and $p_j(m_j)$ the conditional Born probability given preceding outcomes. At a specified schedule, $P_{\rm B}(\mathbf m)=\prod_jp_j(m_j)$, and
\begin{equation}
 P_b^{\rm g}(\mathbf m)=\frac{P_{\rm B}(\mathbf m)^b}{Z_b},\qquad
 P_b^{\rm seq}(\mathbf m)=\prod_j\frac{p_j(m_j)^b}{\sum_ap_j(a)^b}.
 \label{eq:ensembles}
\end{equation}
Here $Z_b=\sum_{\mathbf m}P_{\rm B}(\mathbf m)^b$. These are different history ensembles, with the same local source of bias: unequal Born probabilities are sharpened or flattened by the exponent. The latter is a more physical picture, and we will use it for numerical simulations; but the former is analytically simpler, and we will use it to show the observation-induced ``freezing''.  

The information in a record is quantified by its \emph{surprisal}, $\II=-\log P_{\rm B}(\mathbf m)\geq0$. Writing $\epsilon=b-1$, the global law becomes
\begin{equation}
 P_b^{\rm g}=\frac{P_{\rm B}e^{-\epsilon\II}}{Z_b},\qquad
 Z_b=\E_{\bm m}e^{-\epsilon\II},
 \label{eq:entropy}
\end{equation}
where $\E_{\bm m}$ averages over records drawn with Born probabilities. Positive $\epsilon$ favors less surprising histories; negative $\epsilon$ favors more surprising ones. This is most evident from the relative odds of the two records $a,c$ in \eqref{eq:intro_odds} which can now be written as $e^{-\epsilon(\II_a-\II_c)}$.

Treating $Z_b$ like a partition function gives a free-energy analogue, following the thermodynamic formalism of histories~\cite{LecomteThermodynamic2007}: $\mathcal F(b)=-\log Z_b$, with $\partial_b\mathcal F=\E_b^{\rm g}\II$. For a fixed schedule of unitaries and measurements, the Shannon entropy $\mathcal H_b$ of the globally normalized record distribution satisfies
\begin{equation}
 \frac{\dd\mathcal H_b}{\dd b}=-b\operatorname{Var}_b^{\rm g}(\II).
 \label{eq:recordentropy}
\end{equation}
This means that increasing $b$ removes entropy from this distribution; decreasing it adds entropy. Consequently, the departure from Born statistics also has a nonnegative information cost, which can be quantified by relative entropy (Appendix~\ref{app:histories}). This is the precise sense in which observation acts like Maxwell's demon: there is an entropic cost to observations employing the warped Born distribution, just as there is an entropic cost to the demon sorting gas particles.

Similarly, sequential normalization changes the odds at each event from $p_j(a)/p_j(c)$ to $[p_j(a)/p_j(c)]^b$. For the complete record, the leading change is
\begin{equation}
 \log\frac{P_b^{\rm seq}}{P_{\rm B}}
 =-\epsilon\sum_j[-\log p_j(m_j)-\HH(p_j)]+O(\epsilon^2),
 \label{eq:seqentropy}
\end{equation}
where $\HH(p)=-\sum_mp(m)\log p(m)$ is the conditional Shannon entropy. Each outcome's surprisal is compared with the information expected at that event. For a fixed conditional state, $b>1$ reduces the outcome entropy and $b<1$ increases it. The ensemble-averaged information cost is the sum of the conditional relative-entropy costs along the records. Simple changes to the sequential detector protocol retain this monotonic conditional entropic effect and, with a rule for whole-run success selection, recover the global mapping exactly (Appendix~\ref{app:sequentialprotocols}). We next connect this common preference to particle motion in a globally normalized model where its collective consequences can be found explicitly.

\begin{figure}[t]
\includegraphics[width=\columnwidth]{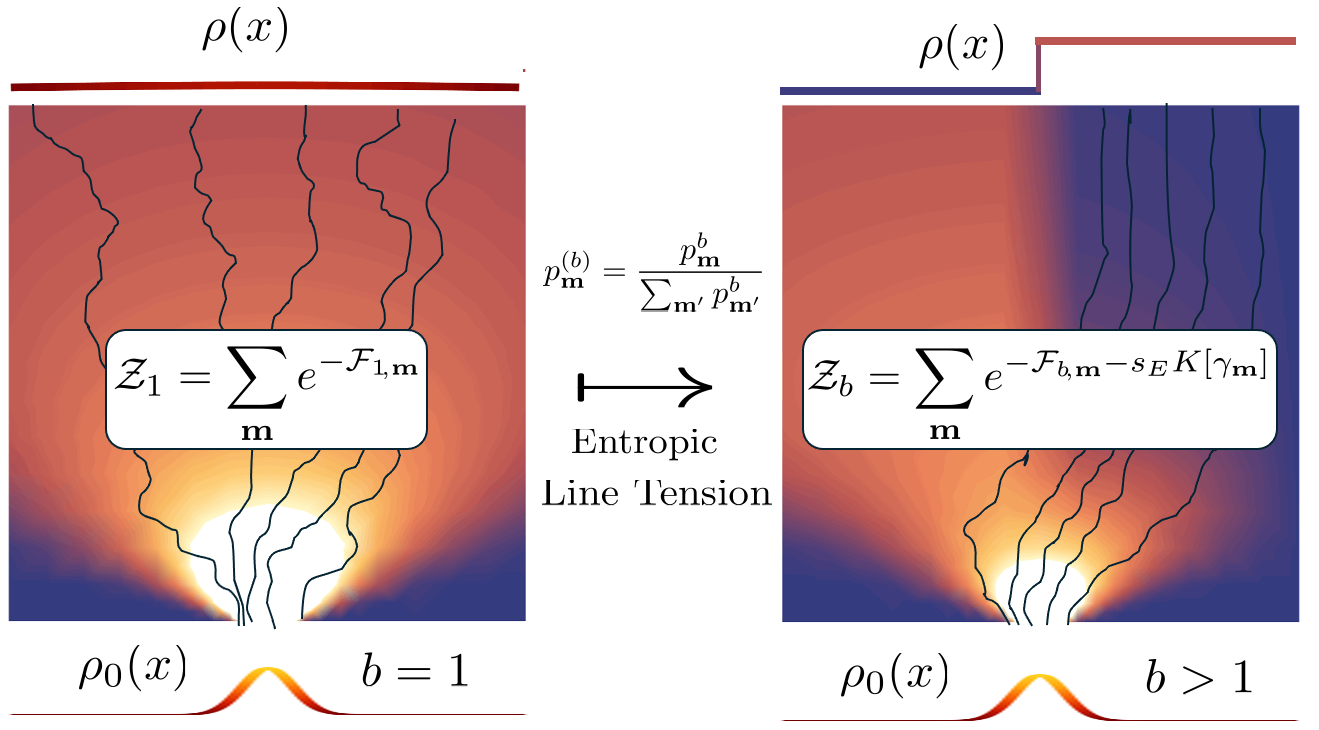}
\caption{Regular diffusion and entropic selection of particle histories. Left: at $b=1$, a localized distribution spreads toward a uniform gas. Right: for $b>1$, globally weighted histories favor fewer uncertain readouts and can select a phase-separated gas. The inset schematically sums over measurement records $\mathbf m$, each with reference cost $\mathcal F_{b,\mathbf m}$ and a collection of particle paths $\gamma_{\mathbf m}$. The total activity $K[\gamma_{\mathbf m}]=\sum_{\gamma\in\gamma_{\mathbf m}}\int dK_\gamma$ counts all particle hops in that record; the extra factor $e^{-\sE K[\gamma_{\mathbf m}]}$, with $\sE>0$, favors less motion.}
\label{fig:biased}
\end{figure}

\textit{A solvable model.---}
To see how this preference can reorganize a gas, consider a particularly simple protocol on a periodic chain. Starting from an occupation configuration, apply a symmetric number-conserving unitary to neighboring sites and measure one endpoint. Conservation fixes the other occupation, so this single-site readout resolves the update entirely. Repeating the procedure gives particles that hop to neighboring vacancies in either direction with equal probability, but cannot share a site. At $b=1$, this is the symmetric exclusion process (SEP), a standard lattice model of diffusion.

Now let $r$ be the Born exchange probability, with $0<r<1$, and write $\alpha=(1-r)^b$, $\beta=r^b$. Their sum $g=\alpha+\beta$ is the total warped weight of an unlike pair's two possible outcomes. Equal neighbors instead have one certain outcome and weight one. Thus $b>1$ gives $g<1$, penalizing repeated uncertain readouts, while $b<1$ rewards them. Global normalization reinforces this difference: a packed region offers fewer uncertain choices than a mixed gas, and thus, for $b > 1$ it is entropically favored at late times without any attractive interaction (Fig.~\ref{fig:biased}).

This connection to motion lets us use an established model of biased diffusion. The \emph{activity} $K$ is simply the number of particle exchanges in a history. An activity-biased SEP weights each ordinary SEP history by $e^{-\sE K}$: positive $\sE$ favors less motion and negative $\sE$ favors more. To reproduce the warped weights exactly, the reference SEP has exchange probability $1-\alpha$, and hence stay probability $\alpha$. Matching its exchange weight $(1-\alpha)e^{-\sE}$ to $\beta$ then gives $P_b^{\rm g}\propto P_{\rm SEP}e^{-\sE K}$, with
\begin{equation}
 \sE=\log\frac{1-\alpha}{\beta}
 =(b-1)\frac{\HH(r)}r+O((b-1)^2),
 \label{eq:tilt}
\end{equation}
where $\HH(r)=-r\log r-(1-r)\log(1-r)$. The ``activity field'' $\sE$ is therefore the strength of this preference, with the same sign as $b-1$.

For a continuous-time version, update each bond at Poisson rate $\nu$ and sum unnormalized weights over schedules and outcomes before normalizing globally. This gives the activity-biased SEP generator exactly. The joint ensemble retains the same surprisal tilt and relative-entropy cost; details of the mapping are given in Appendix~\ref{app:mapping}.

We can now describe the gas at long wavelengths using a density $\rho$ and current $j$. With $x=i/L$ and time $\tau$ measured in diffusion times,\footnote{Here $\tau=kt/L^2$, with reference SEP diffusion constant $k=\nu(1-\alpha)$ in lattice units; $D$ denotes the unmonitored chain's diffusion constant.} conservation requires $\partial_\tau\rho+\partial_xj=0$. Ordinary diffusion favors $j=-\partial_x\rho$, with fluctuations set by the mobility $\sigma(\rho)=2\rho(1-\rho)$. This also counts the local-equilibrium density of particle--vacancy pairs. In contrast, the activity bias becomes a preference among density profiles: writing $\zeta=\sE L^2$, the weak-bias history weight is proportional to $e^{-\mathcal A}$, where~\cite{LecomteGarrahanVanWijland,JackThompsonSollich}
\begin{equation}
 \mathcal A=L\int\dd\tau\dd x\left[
 \frac{(j+\partial_x\rho)^2}{2\sigma(\rho)}+\zeta\sigma(\rho)\right].
 \label{eq:action}
\end{equation}
The first term describes fluctuating diffusion; the second is the entropic bias expressed through opportunities for motion.

\textit{Biased steady states.---}
The consequences of this extra term are striking. Deep inside long globally weighted histories, $b=1$ gives the usual uniform fixed-number equilibrium. For $b>1$, the gas can reduce its activity cost by collecting into a dense region and leaving the remainder empty. A density modulation develops,\footnote{At finite $L$, the uniform profile loses stability at $\sE=\pi^2/[\sigma(\rho_0)L^2]$ to leading hydrodynamic order. This onset vanishes as $L\to\infty$ and is distinct from the measurement-rate crossover below.} becoming a nearly frozen domain with motion confined to its boundaries. Within hydrodynamics, a uniform profile costs order $\zeta$, whereas two interfaces cost only order $\sqrt\zeta$, so segregation wins at large positive $\zeta$ (Appendix~\ref{app:hydro}). The gas has reduced uncertainty in its future readouts by arranging itself so that most updates have nothing to do.

For $b<1$, the same mechanism has the opposite effect. The activity term rewards particle--vacancy pairs and resists long-wavelength density inhomogeneity. Expanding about a uniform profile gives a hyperuniform gas, whose large-scale number fluctuations are suppressed. To distinguish these states, consider the structure factor
\begin{equation}
 S_{\rm occ}(q)=\frac1L\E\left|\sum_j(n_j-\rho_0)e^{-iqj}\right|^2.
 \label{eq:Sqdef}
\end{equation}
In the uniform weak-bias regime, the action gives
\begin{equation}
 S_{\rm occ}(q)\simeq\frac{\chi_0|q|}{\sqrt{q^2-8\sE\chi_0}},
 \qquad\chi_0=\rho_0(1-\rho_0).
 \label{eq:Sq}
\end{equation}
Thus the Born gas ($s_E = 0$) has a nonzero small-$q$ limit, while the hyperuniform gas ($s_E < 0$) has $S_{\rm occ}(q)\to0$. For $s_E > 0$, the uniform state becomes unstable and a macroscopic domain instead gives $S_{\rm occ}(2\pi/L)\propto L$. The domain may wander in a periodic system, restoring a uniform average density, but its long-wavelength correlations still reveal the segregation.

\begin{figure}[t]
% Crop the original panel (c); cover only its obsolete panel letter.
% The original numerical figure and every plotted datum remain unchanged.
\centering
\begingroup
\setlength{\unitlength}{\dimexpr\columnwidth*3/2264\relax}
\begin{picture}(566,655)
\put(0,0){\includegraphics[trim=302.4bp 0 0 0,clip,width=0.75\columnwidth]{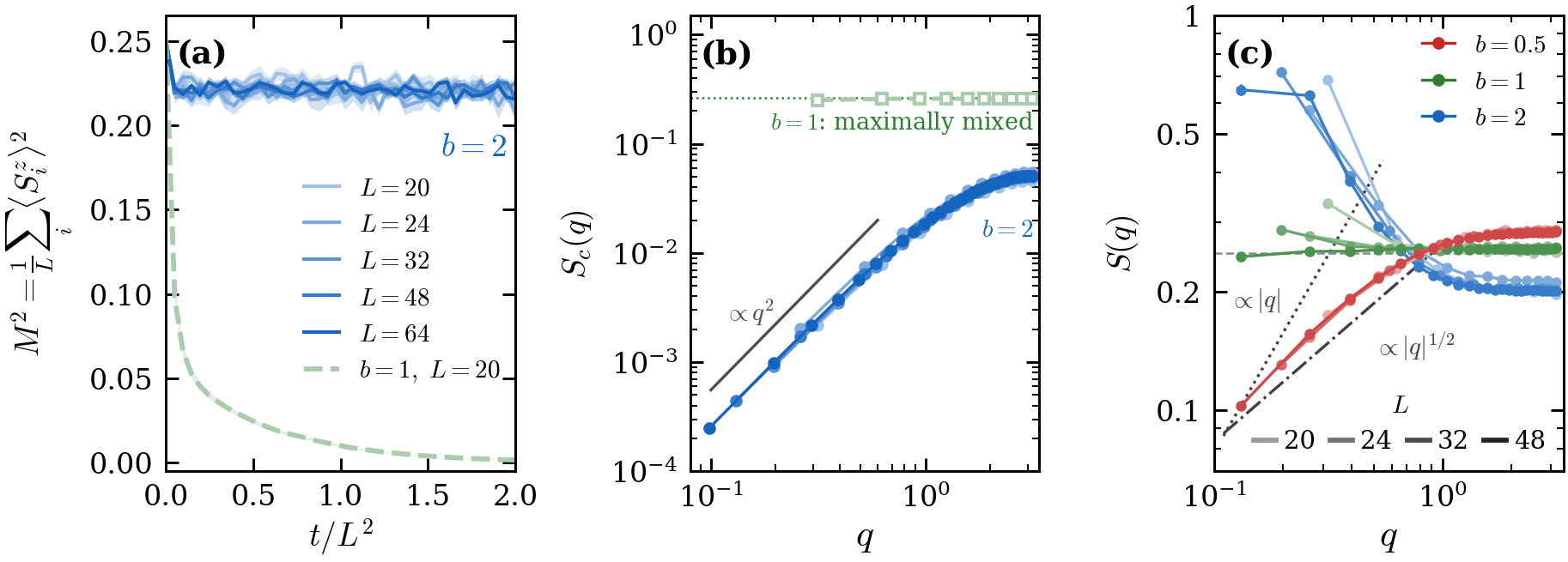}}
\put(163,572){\color{white}\rule{60\unitlength}{46\unitlength}}
\end{picture}
\endgroup
\caption{Numerical signatures of the entropic bias in sequential pure-state histories on open, half-filled chains, $L=20$--48. The full structure factor is enhanced at small $q$ for $b=2$ (blue), nearly flat for $b=1$ (green), and suppressed for $b=0.5$ (red). Spectra average the latter half of each run. Power laws are guides, not fitted asymptotic exponents. Methods and definitions are in Appendix~\ref{app:numerics}.}
\label{fig:magdata}
\end{figure}

\textit{Numerics.---}
These small-$q$ signatures also provide a way to look beyond the solvable protocol. Figure~\ref{fig:magdata} directly samples the sequential measurement histories. Here $S(q)=L^{-1}\E_{\rm seq}\langle\delta n_q^\dagger\delta n_q\rangle$, with $\delta n_q=\sum_j(n_j-1/2)e^{-iqj}$, includes the full density correlations. Relative to the nearly flat Born reference, $b<1$ suppresses long-wavelength fluctuations, with a decreasing small-$q$ spectrum consistent with a hyperuniform regime. For $b>1$, the opposite trend---enhanced low-$q$ weight---indicates clustering. The exact SEP laws $S(q)\sim|q|$ and $S(2\pi/L)\sim L$ need not carry over to this protocol: vanishing fluctuations and growing long-wavelength weight are the qualitative distinctions of interest. The accessible data corroborate analogous clustering and suppression of long-wavelength fluctuations, consistent with the analytic model's phase-separated and hyperuniform regimes. Appendix~\ref{app:numerics}, Fig.~\ref{fig:stationarity}, tests the late-time settling of the longest-wavelength mode.

A complementary history-averaged density-matrix ensemble also retains local polarization for $b>1$. Here branches are summed and weights evaluated on the current ensemble density matrix. Its polarization $M^2=L^{-1}\sum_i\langle S_i^z\rangle^2$, with $S_i^z=n_i-1/2$, remains finite for $b=2$ while relaxing toward zero for $b=1$ (Appendix~\ref{app:numerics}, Fig.~\ref{fig:historyaverage}). This provides complementary evidence of clustering-related inhomogeneity that survives history averaging; $M^2$ alone does not establish a macroscopic domain.

\textit{Consequences for the replica trick.---}
The global convention also enters calculations that do not set out to change Born's rule. Evolving $n$ copies of an unnormalized trajectory density matrix with a common measurement record, then normalizing the contraction by its common-record trace, produces a $2n$-Born-normalized quantity. This structure appears in calculations using generalized entropy moments~\cite{BaoChoiAltmanTheory,LeonticaMcGinley,BaoChoiAltmanSymmetry,BentsenSahuSwingle,FengQuantumTrees,McGinleyHoMalz2025}.

To see the connection, write $\tilde\varrho_{\mathbf m}=P_{\rm B}(\mathbf m)\varrho_{\mathbf m}$ for the unnormalized conditional state. Its $n$-copy product contains $P_{\rm B}^n$. A particular example is the subsystem R\'enyi moment, obtained by a cyclic contraction:
\begin{equation}
 \frac{\sum_{\mathbf m}\operatorname{Tr}(\tilde\varrho_{A,\mathbf m}^{\,n})}
 {\sum_{\mathbf m}(\operatorname{Tr}\tilde\varrho_{\mathbf m})^n}
 =\sum_{\mathbf m}P_{b=n}^{\rm g}(\mathbf m)
 \operatorname{Tr}(\varrho_{A,\mathbf m}^{\,n}).
 \label{eq:replicas}
\end{equation}
For the von Neumann limit, the physical result is recovered through the appropriate $n \to 1$ limit; fixed higher moments require their Born weighting to be treated separately. Circuit averaging retains the history tilt, together with any explicit schedule weights (Appendix~\ref{app:replica}). Integer $n>1$ therefore selects the positive-bias side: a density-ordering saddle may reflect biased histories rather than the qualitative dynamics of a monitored gas with Born-weighted histories.

Whether this preference compounds depends on the time between observations. A mode of wavelength $L$ relaxes on $t_{\rm diff}\sim L^2/D$. Near half-filling the density-memory contribution is suppressed when\footnote{This estimate assumes local renewal and equilibration of conditional probabilities between observations; diffusion of the mean alone is insufficient. Appendix~\ref{app:rate} distinguishes a local waiting time from the total event rate $L\Gamma$.}
\begin{equation}
 \Gamma t_{\rm diff}\lesssim1,\qquad\Gamma\lesssim D/L^2.
 \label{eq:measurementcrossover}
\end{equation}
As diffusion removes the probability differences, the bias from this density-memory contribution becomes small. At exactly half filling, an ideal equilibrated readout has equal probabilities and its odds are unchanged by warping. In other words, equal conditional branch weights remove the outcome bias. Given these conditions, $\Gamma L^2/D$ controls whether the next observation at a site occurs before its density memory relaxes. The scale $\sE L^2$ measures the competition of activity bias with that same relaxation. Both arise from slow particle diffusion, but their quantitative relation depends on the protocol. Forthcoming tower-of-states work numerically identifies a phase boundary at $\Gamma\sim D/L^2$ in its finite-replica model~\cite{OgunnaikeTower}. Thus large-system, long-time limits and the $n\to1$ limit must be controlled together.

\textit{Discussion.---}
We have shown that replacing Born probabilities by $p_b(m)\propto\|P_m\ket\psi\|^{2b}$ introduces an entropic bias through transitions between local occupation measurements. Its sign is set by $b-1$. This effect manifests for the two normalization schemes available to the warped rule.  At fixed schedule, positive warping removes entropy from the global record distribution and negative warping adds it. This can be seen by a ``tilt'' in the relative surprisal, a measure of the unlikeliness of certain histories. Sequential normalization shares this local preference, while its collective dynamics depend on the more complicated intervening coherent evolution.

For a solvable monitored gas model normalizing global trajectories, the two signs of $b-1$ select phase separation and hyperuniformity in the long-time, thermodynamic limit. We corroborate these regimes with numerical tests of two circuits with normalization at each measurement step: one sums histories in a density matrix, while the other samples conditional pure states. They support analogous clustering and hyperuniformity behavior beyond the exact mapping, suggesting that these effects extend beyond the solvable gas. Much as Maxwell's demon turns microscopic sorting into macroscopic order, altered measurement odds can reorganize a monitored gas. These warped measurements can enact a statistical ``cooling,'' suppressing measurement histories with high kinetic activity and biasing towards clustered configurations, while the opposite bias suppresses long-wavelength density fluctuations.

This common-record, normalized linear $n$-copy construction also selects a $2n$-Born rule with consequences for replica calculations involving particle-number symmetry and diffusing particles. For number-conserving systems, fixed-$n>1$ calculations using this construction will generically suffer from the entropic bias, potentially changing the qualitative particle dynamics. The importance of restoring Born weighting through the physical replica limit is explicit in the field theory of particle-number sharpening~\cite{barratt_field_2022}. Forthcoming work on the tower of states describes a particular finite-replica ensemble in which this warped weighting must be accounted for~\cite{OgunnaikeTower}.  Additionally, the bias seen here is related to a number of few-body pathologies outlined by Scott Aaronson~\cite{AaronsonIsland} that are discussed in Appendix~\ref{app:fewbody}. 

A natural future direction would be to study other systems that alter the transition amplitudes between measurements.  In particular, one could consider variants of quantum mechanics involving spontaneous collapse alongside ordinary Born measurements~\cite{BassiCollapse}. Transition amplitudes would depend on the number, locations, and outcomes of intervening collapses; sufficiently rapid particle localization could produce a Zeno-like regime near the initial configuration. Deriving those histories would establish whether they carry an analogous entropic preference.

\textit{Acknowledgements.---} We thank Sam Garratt and Zijian Wang for key insights and Dan Borgnia, Michael Sonner, and Rishi Lohar for useful discussions.

\label{maintext:end}
%\clearpage
\bibliography{pborn_mipt_notes}
\clearpage
\appendix
\setcounter{secnumdepth}{2}
\section{Single-site records and entropic response}
\label{app:histories}
Here we derive the record reweighting in Eqs.~\eqref{eq:entropy} and \eqref{eq:seqentropy}, and quantify its information cost. Keeping the two normalizations separate makes clear which entropy statements apply to complete records and which apply at each observation.

The microscopic detector measures one occupation at a time. For an initial pure state and an external protocol $\omega$ specifying gates and detector events, let $p_j(m\mid\mathbf m_{<j},\omega)$ be the conditional Born probability. The conditional state following a specified record is the same for all $b$: only its sampling weight changes. Detector clocks and any gate randomness have a prescribed law $Q(\dd\omega)$ independent of $b$.

At fixed $\omega$, write $I_j=-\log p_j(m_j)$ and $\II=\sum_jI_j$. The two ensembles obey
\begin{align}
 P_b^{\rm g}&=P_{\rm B}e^{-(b-1)\II}/Z_b,\notag\\
 P_b^{\rm seq}&=P_{\rm B}e^{-(b-1)\JJ_b},\qquad
 \JJ_b=\II-\sum_j\HH_b(p_j),\label{eq:exactseq}\\
 \HH_b(p)&=\frac{\log\sum_mp(m)^b}{1-b}.\notag
\end{align}
Here $\HH_b$ is the R\'enyi entropy; its $b\to1$ limit is the Shannon entropy. Summing the last sequential outcome, and then each preceding outcome, gives $\E_{\rm B}e^{-(b-1)\JJ_b}=1$. The source-free sequential partition function is therefore one, even though individual histories are reweighted.

To expand about $b=1$, define
\begin{equation}
 V(p)=\sum_m p(m)[\log p(m)+\HH(p)]^2.
\end{equation}
Then
\begin{equation}
 \log\frac{P_b^{\rm seq}}{P_{\rm B}}
 =-\epsilon\JJ-\frac{\epsilon^2}{2}\sum_jV(p_j)+O(\epsilon^3).
\end{equation}
Here $\JJ=\sum_j[-\log p_j(m_j)-\HH(p_j)]$ is the first-order record observable in Eq.~\eqref{eq:seqentropy}. The increments of $\JJ$ have zero conditional Born mean and zero covariance at distinct events. Thus
\begin{align}
 \mathcal F_T\equiv\E_{\rm B}\JJ^2&=\E_{\rm B}\sum_jV(p_j),\notag\\
 \E_b^{\rm seq}O-\E_{\rm B}O&=-\epsilon\E_{\rm B}(O\JJ)+O(\epsilon^2),\notag\\
 D(P_b^{\rm seq}\|P_{\rm B})&=\tfrac12\epsilon^2\mathcal F_T+O(\epsilon^3).
\end{align}
$D(P\|Q)=\sum P\log(P/Q)$ is relative entropy. $\mathcal F_T$ is the history Fisher information about $b$ at the Born point. At finite $L,T$ these expansions require the displayed moments to exist. Zero-probability branches are omitted. At site $i$, let $p_i(m)$ be the conditional Born distribution and $w_{b,i}(m)=p_i(m)^b/\sum_a p_i(a)^b$. For stationary Poisson monitoring, the exact chain rule gives the long-time relative-entropy rate
\begin{equation}
 d_b=\Gamma\sum_i\E_{\mu_b}D(w_{b,i}\|p_i),
\end{equation}
where $\mu_b$ is the stationary conditional-state distribution. Under regular near-Born stationary response, $d_b=\epsilon^2 f_{\rm I}/2+o(\epsilon^2)$ with $f_{\rm I}=\lim_{T\to\infty}\mathcal F_T/T$. These assumptions concern finite $L$; a singular thermodynamic limit requires separate control.

\subsection{Entropy change and information cost}
We now establish the entropy response in Eq.~\eqref{eq:recordentropy} and the accumulated relative-entropy cost quoted in the main text.

For a fixed external protocol, differentiating the global partition function gives $\partial_b\mathcal F=\E_b^{\rm g}\II$ and $\partial_b^2\mathcal F=-\operatorname{Var}_b^{\rm g}(\II)$. Since $\log P_b^{\rm g}=-b\II+\mathcal F$, its Shannon entropy is $\mathcal H_b=b\E_b^{\rm g}\II-\mathcal F(b)$, yielding Eq.~\eqref{eq:recordentropy}. These identities hold on the common support of the record distributions, with finite information moments. The relative entropy is $D(P_b^{\rm g}\|P_{\rm B})=\mathcal F-(b-1)\E_b^{\rm g}\II$. Its near-Born expansion is $\epsilon^2\operatorname{Var}_{\rm B}(\II)/2+O(\epsilon^3)$.

For the sequential rule at a fixed pre-measurement state, write $w_b(m)=p(m)^b/\sum_ap(a)^b$. Direct differentiation similarly gives
\begin{equation}
 \frac{\dd\HH(w_b)}{\dd b}=-b\operatorname{Var}_{w_b}(\log p).
\end{equation}
Thus increasing $b$ reduces the conditional outcome entropy. The exact chain rule for the full record cost is
\begin{equation}
 D(P_b^{\rm seq}\|P_{\rm B})
 =\sum_j\E_b^{\rm seq}D(w_{b,j}\|p_j).
\end{equation}
The expectation includes the distribution of preceding records. These preceding-record probabilities change with $b$, so the monotonic conditional entropy at a fixed state does not by itself prove monotonicity of the full sequential record entropy. The conditional entropic preference and the nonnegative accumulated cost are the statements used in the main text.

\subsection{The local density response}
To connect record information to particle motion, we expand the sequential measurement rule near the Born point. The resulting density drift shows how local occupation probabilities and occupation correlations enter the bias.

For conditional occupations $u_i=\langle n_i\rangle$ and connected correlations $C_{ij}=\langle n_in_j\rangle-u_iu_j$, the additional measurement-induced drift is
\begin{equation}
 (\mathcal L_b-\mathcal L_1)\langle n_j\rangle
 =\epsilon\Gamma\sum_i C_{ij}\log\frac{u_i}{1-u_i}+O(\epsilon^2).
\end{equation}
This follows by expanding the binary escort probability $u^b/[u^b+(1-u)^b]$. Deterministic sites contribute zero by continuity. At fixed total particle number, $\sum_jC_{ij}=0$, so the bias conserves particle number. It is not a closed equation for the averaged density: the conditional-state correlations must also be evolved.

\section{Resolved bonds and the global SEP mapping}
\label{app:mapping}
This section derives the exact activity field in Eq.~\eqref{eq:tilt} from a resolved two-site update. It also explains why global and sequential normalization give different exclusion processes in this particular protocol.

Let $C^e$ exchange an unlike pair on bond $e$, and define
$D_e\ket C=(n_i-n_{i+1})^2\ket C$ and
$X_e\ket C=(n_i-n_{i+1})^2\ket{C^e}$. Matrices act on column probability vectors. The unnormalized global transfer and normalized sequential update are
\begin{align}
 T_{b,e}&=\Id-(1-\alpha)D_e+\beta X_e,\notag\\
 S_{b,e}&=\Id+\frac\beta g(X_e-D_e).
\end{align}
To match the active-pair stay weight, introduce a reference SEP whose stay probability is $\alpha$ and whose exchange probability is $1-\alpha$. Multiplying each accepted exchange by $e^{-\sE}$ then gives $\beta$, with $\sE$ as in Eq.~\eqref{eq:tilt}. This auxiliary exchange probability differs from the sequentially normalized probability $\beta/g=1-\alpha/g$. For $0<r<1$, $g<1$ when $b>1$, so $1-\alpha>\beta$ and $\sE>0$; both inequalities reverse when $b<1$.

For a fixed schedule of $M$ updates, let $A$ count unlike-pair opportunities and $K$ count accepted exchanges. Multiplication of elementary weights gives
\begin{align}
 P_{\rm B}^b&=P_b^{\rm seq}g^A
 =P_{\rm SEP,\,1-\alpha}e^{-\sE K},\notag\\
 Z_b&=\bra{\mathbf1}T_{b,e_M}\cdots T_{b,e_1}\ket{C_0}.
\end{align}
Both expressions represent the same global ensemble. One uses sequential SEP as its reference and weights active opportunities; the other uses the auxiliary probability $1-\alpha$ and weights accepted exchanges. Neither identifies global normalization with the physical sequential law. In fact
\begin{equation}
 P_b^{\rm seq}=P_{\rm SEP,\,1-\alpha}e^{-\sE K-A\log g}.
\end{equation}
The second term compensates the pure activity tilt. Unless $A$ is record independent, dropping it changes the ensemble. The sequential matrix preserves the uniform fixed-number measure.

For a global conditional probability at update $j$, let $h_j(C)$ be the total weight of all later trajectories from $C$. Then
\begin{equation}
 P_b^{\rm g}(C'|C,j)=\frac{(T_{b,e_j})_{C',C}h_j(C')}{h_{j-1}(C)},
 \qquad h_M=1.
\end{equation}
For an active bond the exchange probability is
\begin{equation}
 \frac{\beta h_j(C^e)}{\alpha h_j(C)+\beta h_j(C^e)}.
\end{equation} It reduces to $\beta/g$ only when both measurement trajectories have equal weight. A scalar global normalization therefore cancels from complete-record odds without eliminating future-dependent conditioning.

\subsection{Annealed clocks and a deterministic alternative}
The continuous-time mapping requires a choice of how to average the update schedule. We first derive the generator obtained by summing unnormalized histories over Poisson clocks, then give a deterministic small-step limit with the same generator.

For fixed protocol normalization, the joint law is
$Q(\dd\omega)P_{\rm B}(\mathbf m|\omega)^b/Z_b(\omega)$.
Annealed normalization instead gives
\begin{equation}
 \mathbb P_b^{\rm ann}(\dd\omega,\mathbf m)
 =\frac{Q(\dd\omega)P_{\rm B}(\mathbf m|\omega)^b}{\mathcal Z_b},
 \quad\mathcal Z_b=\E_QZ_b(\omega).
\end{equation}
Relative to $\mathbb P_{\rm B}=Q P_{\rm B}$, its likelihood ratio is
$e^{-\epsilon\II}/\mathcal Z_b$, and
$\mathcal Z_b=\E_{Q,\rm B}e^{-\epsilon\II}$.
Thus it has the same entropic coupling, while its protocol marginal is $QZ_b/\mathcal Z_b$. In particular, for a $b$-independent observable,
\begin{align}
 \delta\E^{\rm ann}O&=-\epsilon\operatorname{Cov}_{Q,\rm B}(O,\II)+O(\epsilon^2),\notag\\
 \delta\E^{\rm fixed}O&=-\epsilon\E_Q\operatorname{Cov}_{\rm B|\omega}(O,\II)+O(\epsilon^2).
\end{align}
The two first-order responses differ by the covariance of the protocol-conditional means. The clocks are not themselves raised to a power; their relative statistical weights change through the outcome partition function.

The joint ensemble retains the same surprisal tilt and relative-entropy cost: $D(\mathbb P_b^{\rm ann}\|\mathbb P_{\rm B})=-\log\mathcal Z_b-(b-1)\E_b^{\rm ann}\II$. The fixed-schedule Shannon theorem does not transfer unchanged to its joint entropy. For a discrete protocol law $Q$, that entropy additionally contains $-\E_b^{\rm ann}\log Q$; its derivative need not have a fixed sign. For continuous schedules the relative-entropy statement avoids any choice of reference measure for a differential entropy.

For independent bond attempts at rate $\nu$, a short-time annealed transfer is
$\Id+\nu\dd t\sum_e(T_{b,e}-\Id)+O(\dd t^2)$.
The exact generator, including the escape weights, is
\begin{equation}
 W_s=k\sum_e(e^{-s}X_e-D_e),\quad k=\nu(1-\alpha),\quad s=\sE.
 \label{eq:generator}
\end{equation}
Thus $k$ sets the reference diffusion constant in lattice units. By contrast, sequential averaging yields $W_0(\nu\beta/g)$. The finite-gate Poisson construction assumes instantaneous gates followed by one endpoint measurement; it is distinct from independent detectors acting during overlapping coherent dynamics.

For a deterministic ordered cycle, take $r\to0$, set its duration $\Delta=(1-\alpha)/k$, and tune $b=b(r)$ to satisfy $\beta/(1-\alpha)=e^{-s}$. At fixed $L,T$, the cycle transfer is $\Id+\Delta W_s+O(\Delta^2)$, and its repeated product tends to $e^{TW_s}$. The required scaling is
\begin{equation}
 b(r)=1+\frac{s}{\log(1/r)+1}+O\!\left(\log^{-2}(1/r)\right).
\end{equation}
At fixed $b\ne1$ the field instead diverges as
$\sE=(b-1)\log(1/r)+\log b+O(r)$.
For coherent pulses with $r=\sin^2(J\tau)$, a finite transport rate requires a slow-time rescaling or $J\propto\tau^{-1/2}$ with $J\tau\to0$. The microscopic limit is taken before the thermodynamic and long-record limits.

\subsection{A diffusive control}
A useful control retains diffusion while removing uncertain readouts. It isolates the role of amplitude mixing relative to the local measurement basis.

Restrict the gates to occupation-basis permutations and phases, and initialize a basis configuration. Every occupation readout is then deterministic, so both warped schemes have the same histories for every $b$. Random externally chosen nearest-neighbor swaps still produce ordinary diffusion. This control isolates mixing relative to the operational basis as a necessary ingredient of the measurement bias; changing the exponent alone is not a theorem of collective ordering.

\section{Modified sequential protocols}
\label{app:sequentialprotocols}
Here we give two explicit ways to make the sequential rule sensitive to the local particle environment. Resolving an additional detector bit yields an exactly solvable preference for fewer or more particle--vacancy interfaces. A separate success-selection procedure cancels the sequential normalizers and reproduces the global history law, including its collective phases. The first construction changes the instrument; the second changes which runs are retained.

\subsection{A detector bit generates an interface interaction}
Consider resolved configurations on a periodic chain at fixed particle number, and let
\begin{equation}
 A(C)=\sum_i(n_i-n_{i+1})^2
\end{equation}
count particle--vacancy interfaces. At each bond attempt, a balanced number-conserving gate produces the two unlike-pair orientations with equal Born probabilities. A readout of one endpoint distinguishes them. If exchanging the pair changes $A$, its change is $\pm2$; denote the candidates by $C_{\rm lo}$ and $C_{\rm hi}$. Which has more interfaces depends only on the two neighboring occupations, already known in this resolved protocol.

We now refine the endpoint detector: the high-interface outcome carries a noise bit with probabilities $1-\eta,\eta$, with $0<\eta<1$. The resolved joint outcomes have Born weights
\begin{equation}
 p({\rm lo})=\tfrac12,\qquad
 p({\rm hi},0)=\tfrac{1-\eta}{2},\qquad
 p({\rm hi},1)=\tfrac\eta2.
 \label{eq:detectorbit}
\end{equation}
The occupation and bit are resolved as one instrument before applying the power $b$. Adding a bit after choosing a warped occupation outcome would have no effect on that choice. Thus this construction enlarges the single-site detector's outcome alphabet; it is not the original binary projective rule. We apply the warped rule to the fully resolved Kraus outcomes, with the detector setting chosen from the already known neighboring occupations.

For fixed neighboring occupations, an explicit instrument on the unlike-pair block is
\begin{align}
 K_{\rm lo}&=P_{\rm lo}U,\notag\\
 K_{{\rm hi},0}&=\sqrt{1-\eta}\,P_{\rm hi}U,\qquad
 K_{{\rm hi},1}=\sqrt\eta\,P_{\rm hi}U,
 \label{eq:bitkraus}
\end{align}
where $U$ is the balanced gate and $P_{\rm lo},P_{\rm hi}$ are its endpoint occupation projectors restricted to that block. For equal-interface candidates use an ordinary endpoint readout; equal occupations remain unchanged. Taking the direct sum over these blocks and the spectator occupations defines a complete, number-conserving instrument. Its Born channel is unital: both $\sum_aK_a^\dagger K_a$ and $\sum_aK_aK_a^\dagger$ are the identity. Every resolved input remains resolved, and forgetting the bit at $b=1$ leaves ordinary symmetric exclusion.

Sequential normalization of Eq.~\eqref{eq:detectorbit}, followed by forgetting the bit, gives
\begin{align}
 \frac{P_b({\rm hi})}{P_b({\rm lo})}
 &= (1-\eta)^b+\eta^b=e^{-2J_b},\notag\\
 J_b&=-\frac12\log[(1-\eta)^b+\eta^b]\notag\\
 &=\frac{b-1}{2}\HH(\eta)+O((b-1)^2).
 \label{eq:interfacecoupling}
\end{align}
The entropy of the extra detector bit therefore sets the interaction strength near the Born point. Both input orientations have the same pair of candidates, so the exchange probability on an unlike pair is
\begin{equation}
 w_b(C\to C^e)=\frac1{1+e^{J_b[A(C^e)-A(C)]}}.
\end{equation}
These are particle-conserving heat-bath (Kawasaki) updates. Their forward-to-reverse ratio is $e^{-J_b[A(C^e)-A(C)]}$, establishing detailed balance with
\begin{equation}
 \pi_b(C)=\frac{e^{-J_b A(C)}}{\mathcal Z_N(J_b)},\qquad
 \mathcal Z_N(J)=\sum_{C:\,\sum_i n_i=N}e^{-JA(C)}.
 \label{eq:interfacegibbs}
\end{equation}
For positive, configuration-independent bond-attempt rates and finite $b,\eta$, exchanges connect the fixed-number sector, making this the unique stationary law. Since $\partial_J\langle A\rangle=-\operatorname{Var}(A)$ and $\partial_bJ_b>0$, increasing $b$ monotonically suppresses interfaces. This is a density-ordering consequence of the conditional entropic bias, not a monotonicity theorem for the entropy of complete sequential records.

The stationary correlations can be obtained exactly. At half filling, set $s_i=2n_i-1$; then $A=(L-\sum_i s_is_{i+1})/2$. Equation~\eqref{eq:interfacegibbs} is the one-dimensional Ising measure with coupling $J_b/2$ and fixed zero magnetization. In the thermodynamic limit, at nonzero wave number,
\begin{equation}
 S(q)=\frac14\frac{1-t_b^2}{1-2t_b\cos q+t_b^2},\qquad
 t_b=\tanh(J_b/2),
 \label{eq:bitspectrum}
\end{equation}
and hence $\lim_{q\to0}S(q)=e^{J_b}/4$, taking $L\to\infty$ first. Positive warping gives positive spatial correlations and clustering; negative warping suppresses long-wavelength fluctuations. This limit remains strictly positive at finite parameters, so the detector-bit construction alone gives neither thermodynamic hyperuniformity nor macroscopic phase separation in one dimension. It exhibits the two trends without identifying its path law with the global activity-biased SEP.

\subsection{Success selection recovers the global history law}
An exact identification with the global protocol is possible by retaining only fully successful sequential runs. This construction keeps the original single-site occupation projectors. The added success flag is sampled externally, with the stated probability, rather than subjected to another warped measurement.

At event $j$, let $Z_j=\sum_a p_j(a)^b$ for its binary Born probabilities. Independently of which outcome is drawn, accept this step with probability $c_b Z_j$, where
\begin{equation}
 c_b=\min(1,2^{b-1}).
\end{equation}
The binary bound $Z_j\leq\max(1,2^{1-b})$ ensures a valid acceptance probability. Discard the whole run upon any failure. At a fixed schedule containing $M$ events,
\begin{align}
 P(\mathbf m,\mathrm{success})
 &=\prod_j\left[c_bZ_j\frac{p_j(m_j)^b}{Z_j}\right]
 =c_b^M P_{\rm B}(\mathbf m)^b,\notag\\
 P(\mathbf m\mid\mathrm{success})
 &=\frac{P_{\rm B}(\mathbf m)^b}{Z_b},\qquad
 P(\mathrm{success})=c_b^M Z_b.
 \label{eq:heraldedidentity}
\end{align}
Thus selection cancels every local normalizer exactly. At fixed schedule the retained law inherits both the global surprisal tilt and its monotonic record entropy, Eq.~\eqref{eq:recordentropy}. The protocol requires knowledge of the conditional Born probabilities; it is a specified sampling procedure, not a claim that a passive binary detector performs this acceptance test automatically.

For the resolved-bond model, the acceptance probability is $c_b g$ on a particle--vacancy pair and $c_b$ on equal neighbors. It therefore depends on the local density pattern. At $b>1$, uncertain active bonds are less likely to survive selection; at $b<1$, they survive more readily than deterministic bonds. This is the missing configuration-dependent weight that ordinary sequential normalization removes.

Poisson schedules require a clock correction because their event count fluctuates. To recover the globally annealed protocol with attempt rate $\nu$ on each bond, use rate $\nu/c_b$ in the selected sequential procedure. Its unnormalized successful-event transfer is $c_b T_{b,e}$, giving
\begin{align}
 W_{\rm keep}
 &=\frac\nu{c_b}\sum_e(c_bT_{b,e}-\Id)\notag\\
 &=W_{\sE}+L\nu(1-c_b^{-1})\Id.
 \label{eq:heraldedgenerator}
\end{align}
Here $W_{\sE}$ is Eq.~\eqref{eq:generator}. The scalar shift cancels from normalized histories at fixed duration $T$. Equivalently, the Poisson schedule likelihood ratio contributes $c_b^{-M}e^{L\nu(1-c_b^{-1})T}$, canceling $c_b^M$ in Eq.~\eqref{eq:heraldedidentity}. The total success probability is $e^{L\nu(1-c_b^{-1})T}\mathcal Z_b(T)$, where $\mathcal Z_b$ is the original annealed partition function. Success can be exponentially rare; its cost is part of the construction.

Consequently, the temporal bulk of long successful runs has exactly the global model's phase-separated regime for $b>1$ and hyperuniform regime for $b<1$, with the limits and structure factors derived in Appendix~\ref{app:hydro}. This establishes an implementation of those phases by a modified, selected sequential procedure. It does not identify unselected runs, or the numerical protocol in Fig.~\ref{fig:magdata}, with the global ensemble.

\section{Particle number conservation and a parity-only comparison}
\label{app:conservation}
Here we separate the general record-information tilt from its density-ordering mechanism. A parity-preserving example shows why a global discrete symmetry need not supply the diffusive density field used in the main text.

Conservation of $\sum_i n_i$ produces both the long-lived density field and the configuration-dependent number of active bonds in the exclusion mapping. A global discrete symmetry alone does not imply this continuity equation. It also does not prevent an entropic tilt of measurement records: Eqs.~\eqref{eq:ensembles}--\eqref{eq:entropy} apply to arbitrary conditional probabilities.

A simple model shows precisely how the density-dependent bias can disappear. On each chosen bond apply $U_\theta=\exp(-i\theta X_iX_{i+1})$, where $X_i$ flips the occupation. This gate preserves global $Z$ parity but not total particle number. Starting from a basis configuration, its two branches leave both occupations unchanged or flip both, with Born probabilities $1-r$ and $r$, where $r=\sin^2\theta$. A measurement of either endpoint resolves these branches, including for equal occupations. All observations are single-site readouts.

The warped transfer is $T_{b,e}=\alpha\Id+\beta F_e$, where $F_e$ flips the two occupations. Every column has sum $g=\alpha+\beta$, independently of the configuration. For a fixed schedule of $M$ updates, $Z_b=g^M$ and
\begin{equation}
 P_b^{\rm g}=P_b^{\rm seq},\qquad
 S_{b,e}=\frac\alpha g\Id+\frac\beta g F_e.
\end{equation}
Warping changes the flip probability but introduces no preference for one density profile over another. On a connected ring, the invariant distribution is uniform within each parity sector. The factor corresponding to an activity potential is a configuration-independent constant. This cancellation relies on the equal branching structure of this example; it is not a theorem that all nonconserving or parity-preserving circuits have no record bias or collective transitions.

\section{Hydrodynamics and collective phases}
\label{app:hydro}
We derive the hydrodynamic action in Eq.~\eqref{eq:action} and use it to obtain the two collective regimes. We first specify the long-time ensemble to which the steady-state correlations refer.

At finite $L$, let $h_s$ be the positive leading eigenvector of the symmetric $W_s$. A configuration at time $t$ in a record of duration $T$ has unnormalized weight
$[e^{tW_s}]_{C,C_0}[\bra{\mathbf1}e^{(T-t)W_s}]_C$.
When $t$ and $T-t$ both exceed the inverse spectral gap, its normalized law becomes $h_s(C)^2/\sum_{C'}h_s(C')^2$. Its generating rate is the leading eigenvalue $\psi_L(s)=\lim_{T\to\infty}T^{-1}\log Z_s$. The long-record limit precedes the thermodynamic limit.

The exact identity
\begin{equation}
 W_s(k)=W_0(ke^{-s})-k(1-e^{-s})\sum_eD_e
\end{equation}
rewrites the exchange tilt as a potential on active bonds. Local equilibrium gives an active-bond density $\sigma(\rho)=2\rho(1-\rho)$. With $x=i/L$, $\tau=kt/L^2$, and $s=\zeta/L^2$, this potential contributes $L\zeta\int\dd x\dd\tau\,\sigma$. The change in reference diffusivity is subleading at fixed $\zeta$. Ordinary SEP fluctuations obey
\begin{equation}
 \partial_\tau\rho+\partial_xj=0,\qquad
 j=-\partial_x\rho+\sqrt{\sigma(\rho)/L}\,\xi,
\end{equation}
where $\xi$ is unit space-time white noise. Its Gaussian weight gives the first term in Eq.~\eqref{eq:action}. This is a controlled leading hydrodynamic description for smooth profiles, not an identity equating hop count with squared net current.

\subsection{Instability and dense domains}
For positive activity field, a density modulation can lower the history cost. We locate its onset and show why a single dense domain is favored deeper in the inactive regime.

Writing $f(\rho)=\rho\log\rho+(1-\rho)\log(1-\rho)$ gives $f''=2/\sigma$. Continuity and periodic boundaries imply
$\int j\partial_x\rho/\sigma\,\dd x=\frac12\partial_\tau\int f(\rho)\dd x$.
After discarding this temporal boundary contribution to the long-time rate, the nonnegative current cost is minimized by a stationary profile with $j=0$. Hence
\begin{equation}
 \psi_L(\zeta/L^2)\simeq-\frac{k}{L}\min_\rho\int_0^1\dd x
 \left[\frac{(\partial_x\rho)^2}{2\sigma(\rho)}+\zeta\sigma(\rho)\right],
\end{equation}
subject to $\int\rho\dd x=\rho_0$. For $\rho=\rho_0+u\cos(2\pi x)$ the quadratic cost is $(\pi^2/\sigma_0-\zeta)u^2$, so the uniform stationary profile becomes unstable at $s_c=\pi^2/(\sigma_0L^2)$. This threshold in the field of the resolved exclusion model is distinct from the observation-versus-equilibration crossover in Eq.~\eqref{eq:measurementcrossover}. Optimizing the induced second harmonic gives a positive quartic coefficient $\pi^2/(4\sigma_0^3)$, so the emerging modulation is continuous. The full stationary SEP solution places the inactive transition at this scale~\cite{LecomteGarrahanVanWijland}.

Deeper in the inactive regime, $1\ll\zeta\ll L^2$, an isolated interface satisfies
\begin{equation}
 \partial_x\rho=\sqrt{8\zeta}\rho(1-\rho),\qquad
 \rho(x)=\frac1{1+e^{-\sqrt{8\zeta}(x-x_0)}}.
\end{equation}
Its cost is $\sqrt{2\zeta}$. A ring has two walls, with dense and dilute lengths set by the conserved mass. The resulting cost $2\sqrt{2\zeta}$ beats the uniform $\zeta\sigma_0$; additional walls cost more. This is macroscopic phase separation, with interface width $\xi=(8s)^{-1/2}$ in lattice sites. The regime requires $1\ll\xi\ll\min(\rho_0L,(1-\rho_0)L)$. Translating the domain restores a uniform mean density but leaves
$S_{\rm occ}(2\pi/L)\simeq L\sin^2(\pi\rho_0)/\pi^2$.

\subsection{Hyperuniform fluctuations}
For negative activity field, the uniform density remains stable but its long-wavelength fluctuations are suppressed. Expanding the action gives the structure factor in Eq.~\eqref{eq:Sq}.

For $\rho=\rho_0+\varphi$, eliminate the current using continuity. In scaled Fourier variables $Q=2\pi n\ne0$ and frequency $\Omega$, the inverse Gaussian covariance is proportional to
\begin{equation}
 \frac{\Omega^2}{\sigma_0Q^2}+\frac{Q^2}{\sigma_0}+\zeta\sigma_0'',
 \qquad \sigma_0''=-4.
\end{equation}
Frequency integration gives
\begin{equation}
 L\langle|\varphi_Q|^2\rangle=\frac{\sigma_0|Q|}{2\sqrt{Q^2+\zeta\sigma_0\sigma_0''}},
\end{equation}
In physical wave number $q=Q/L$, this gives
\begin{equation}
 S_{\rm occ}(q)\simeq\frac{\chi_0|q|}{\sqrt{q^2-8s\chi_0}},\qquad\chi_0=\rho_0(1-\rho_0).
\end{equation}
This formula describes the stable uniform branch, including $s<0$. For negative weak bias,
\begin{equation}
 S_{\rm occ}(q)\sim\sqrt{\frac{\chi_0}{8|s|}}|q|,\qquad
 L^{-1}\ll|q|\ll\sqrt{8|s|\chi_0}\ll1.
\end{equation}
Fixed total particle number removes $q=0$ in every ensemble; hyperuniformity concerns small nonzero wave numbers. The scale separation requires $L\sqrt{|s|}\gg1$; bounded $sL^2$ gives a finite-size crossover. A segment of length $\ell$ beyond the crossover has number variance $2c_s\log\ell/\pi+O(1)$ when $S_{\rm occ}(q)\sim c_s|q|$. At half filling, the XXZ mapping gives the exact thermodynamic slope $c_s=[4\arccos(e^s)]^{-1}$ for all fixed $s<0$~\cite{VanicatMapping}. The small-$|s|$ limit agrees with the Gaussian result. These correlations refer to the temporal bulk of the globally weighted histories.

\section{Observation rate and diffusive memory}
\label{app:rate}
Here we explain the density-memory crossover in Eq.~\eqref{eq:measurementcrossover} through the persistence of local occupation information between observations. The argument explains the scale of the phase boundary numerically identified in the finite-replica model of forthcoming tower-of-states work~\cite{OgunnaikeTower}; it does not independently determine that boundary or its critical behavior. We assume that unmonitored long-wavelength density modes relax diffusively, that local probabilities equilibrate within a fixed-number sector after this memory decays, and that a local renewal description captures the relevant waiting time. Short-time scrambling is taken faster than the longest density relaxation. We illustrate the cancellation at half filling, where an equilibrated single-site measurement has probabilities $(1/2,1/2)$.

For a mode with wave number $q$, let its contribution to a local occupation probability be $u_i(t)-1/2=a_q v_q(i)e^{-Dq^2t}$. Near local equilibrium the binary sequential rule gives
\begin{equation}
 \frac{u_i^b}{u_i^b+(1-u_i)^b}-u_i
 =\epsilon(u_i-\tfrac12)+O\big(\epsilon(u_i-\tfrac12)^3,\epsilon^2\big).
\end{equation}
The log-odds shift is $\epsilon\log[u_i/(1-u_i)]$, or $4\epsilon(u_i-1/2)$ to leading order in the density perturbation.
The bias sensitive to this mode therefore decays with its diffusive memory. Averaging that memory over the waiting time to the next observation at this site yields
\begin{equation}
 \int_0^\infty\dd t\,\Gamma e^{-\Gamma t}e^{-Dq^2t}
 =\frac{\Gamma}{\Gamma+Dq^2}.
 \label{eq:memoryfactor}
\end{equation}
At $q_{\min}=2\pi/L$ on a ring, the crossover occurs at $\Gamma\sim Dq_{\min}^2$, or $\Gamma L^2/D=O(1)$ up to a geometry-dependent constant. It is smooth: for any nonzero $\Gamma$ there is a small probability of a short waiting interval. The rate at which the biased observations accumulate carries an additional factor $\Gamma$, and the deformation supplies $\epsilon$. Equation~\eqref{eq:memoryfactor} motivates the boundary scale but does not fix its critical prefactor. For an exponential local waiting time, the probability of an uninterrupted interval lasting at least $t_{\rm diff}$ is $e^{-\Gamma t_{\rm diff}}$; thus $\Gamma L^2/D$ controls a bridge probability rather than itself being one. The memory factor instead measures how much of a diffusive mode remains when the next readout occurs. In the resolved global model, the comparison of activity bias with the same $L^{-2}$ relaxation scale gives $\sE L^2$. The former scale sets the opportunity to retain and reweight occupation information, the latter its accumulated dynamical effect; neither alone supplies a universal relation between $\Gamma$ and $\sE$.

In the ideal local-equilibrium reset limit at half filling, each subsequent single-site outcome has probability $1/2$ independently of the previous record. For a fixed schedule with $M$ events, $P_{\rm B}(\mathbf m)=2^{-M}$ for every record. Global normalization and sequential normalization then both restore precisely the uniform outcome distribution for every $b$: there is no relative outcome bias to compound. Annealing schedules in the global ensemble can still reweight their event count through $Z_b=2^{M(1-b)}$; this alone is not a preference between equally likely outcomes at a given schedule. Away from half filling, even an equilibrated single-site marginal has unequal outcome weights; the cancellation invoked in the main text is specifically the symmetric case.

These assumptions are stronger than diffusion of the mean density. A pure wavefunction can have fluctuating configuration amplitudes even after its coarse density has equilibrated. Moreover, independent site clocks produce events somewhere in the chain at total rate $L\Gamma$. Equation~\eqref{eq:memoryfactor} uses the waiting time of a local observation, not a globally measurement-free interval. Measurements elsewhere can retain or regenerate correlated information; reducing those effects to the local renewal estimate requires validation in the full monitored model. The same applies when gate speeds, measurement backaction, or additional slow modes alter equilibration. Thus the rate estimate motivates the observed boundary scale in the cited model, while the exact record-information bias remains defined at every measurement rate.

\section{Numerical protocols and observables}
\label{app:numerics}
\begin{figure*}[t]
% Original panels (a,b), cropped without altering their plotted content.
\includegraphics[trim=0 0 138.24bp 0,clip,width=0.9\textwidth]{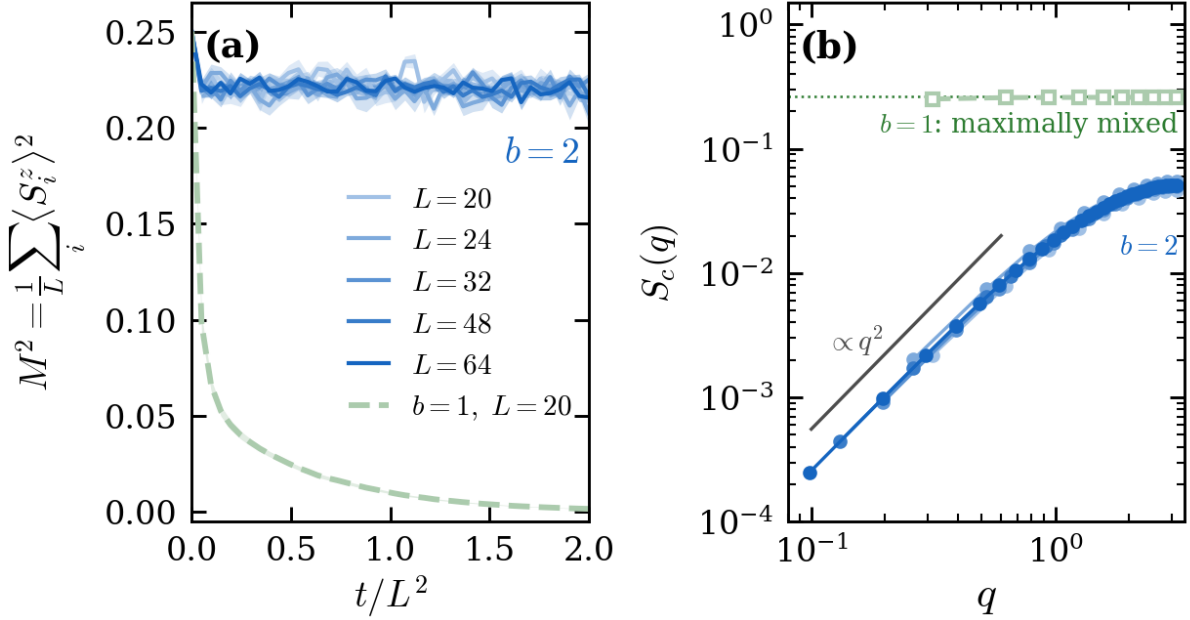}
\caption{Complementary signatures in the history-averaged density-matrix ensemble with Haar-random number-conserving gates. (a) $M^2=L^{-1}\sum_i\langle S_i^z\rangle^2$ remains finite for $b=2$ ($L=20$--64) but relaxes toward zero for $b=1$ ($L=20$). (b) Connected fluctuations $S_c(q)$ about each mixed-state density profile are suppressed for $b=2$; the dotted line is the fixed-number equilibrium value $L/[4(L-1)]$. These weights are evaluated on the ensemble density matrix and differ from the conditional-state weights of Fig.~\ref{fig:magdata}.}
\label{fig:historyaverage}
\end{figure*}

This section specifies the sequential ensemble in Fig.~\ref{fig:magdata} and the complementary history-averaged ensemble in Fig.~\ref{fig:historyaverage}, together with their observables. In particular, it distinguishes order visible after summing histories from correlations evaluated within individual histories.

Both figures use open spin-$1/2$ chains with $S_i^z=n_i-1/2$ and fixed total particle number $N=L/2$. For each size, all realizations and exponents start from the same random product configuration, generated with seed 2026. Odd and even layers address alternating nearest-neighbor bonds. Each eligible bond receives a number-conserving gate with probability $f_U$. The Fourier samples are $q=2\pi n/L$, $n=1,\ldots,L/2$; for these open chains they define the plotted transform rather than periodic-boundary normal modes. Spectra are time-averaged over the latter half of each run, then averaged over independent circuits or trajectories. Error bars and bands show one standard error of the latter average; where smaller than the plotted symbols or line widths, they are not visually resolved.

\subsection{History-averaged density-matrix evolution}
Figure~\ref{fig:historyaverage} sums measurement branches in the density matrix at every step. We give both the update rule and its explicit expansion over histories, so that the meaning of this ensemble average is unambiguous.

The gate has a Haar-random $U(2)$ block on $\{|10\rangle,|01\rangle\}$ and independent uniform phases on $|00\rangle,|11\rangle$, with $f_U=0.3$. After each layer, sites are visited from left to right and the current density matrix is updated by
\begin{equation}
 \rho\longmapsto(1-f_{\rm m})\rho+
 f_{\rm m}\frac{\sum_m p_i(m)^{b-1}P_{i,m}\rho P_{i,m}}
 {\sum_m p_i(m)^b},
 \label{eq:nonlinear_map}
\end{equation}
with $f_{\rm m}=0.6$ and $p_i(m)=\operatorname{Tr}(P_{i,m}\rho)$. A zero-probability branch contributes zero. This is an explicit history sum with normalization at the ensemble level. To see this, decompose the density matrix as $\rho=\sum_h R_h$, where each positive, unnormalized $R_h$ carries a preceding history $h$ and $\sum_h\operatorname{Tr}R_h=1$. Define $g_i(\rho)=\sum_m p_i(m)^b$. Extend each history by either no observation ($\varnothing$) or outcome $m$:
\begin{align}
 R'_{h,\varnothing}&=(1-f_{\rm m})R_h,\\
 R'_{h,m}&=\frac{f_{\rm m}p_i(m)^{b-1}}{g_i(\rho)}
 P_{i,m}R_hP_{i,m}.
 \label{eq:mixed_histories}
\end{align}
Summing these branches reproduces Eq.~\eqref{eq:nonlinear_map}. Their nonnegative traces sum to one and give the history weights; dividing each nonzero branch by its trace gives its conditional state. Unitaries act on every branch before the next update. Repeating this construction therefore retains a sum over complete histories, including measurement locations and outcomes.

The coefficients in Eq.~\eqref{eq:mixed_histories} use the current ensemble probabilities $p_i(m)$, not the probabilities of each normalized branch $R_h/\operatorname{Tr}R_h$. At $b\ne1$ this can also reweight the relative contributions of earlier histories. Figure~\ref{fig:magdata}, by contrast, normalizes the outcome probabilities separately on each conditional pure state. Both are history ensembles, but their weights generally differ.

There are 16 independent circuit realizations at each size and exponent: $b=2$ uses $L=20,24,32,48,64$, and the $b=1$ reference uses $L=20$, with the same circuits at that size. Runs have $T=2L^2$ layers, with measurements of observables at the endpoints and 40 equally spaced intermediate times. With $\langle O\rangle=\operatorname{Tr}(\rho O)$, the plotted observables are
\begin{align}
 M^2(t)&=\E_{\rm circ}\frac1L\sum_i\langle S_i^z\rangle^2,\\
 S_c(q)&=\E_{\rm circ}\overline{\frac1L\sum_{x,y}e^{iq(x-y)}
 [\langle S_x^zS_y^z\rangle-\langle S_x^z\rangle\langle S_y^z\rangle]}.
\end{align}
The overline denotes the late-time average. The brackets first average measurement histories within each circuit; the square in $M^2$ is taken next, before averaging over circuits. Thus $M^2$ detects polarization that survives the history average, rather than polarization only present in individual conditional states. A product state has $S_c(q)=0$, whereas the maximally mixed fixed-number state has $S_c(q)=L/[4(L-1)]$ at the sampled nonzero wave numbers. A finite $M^2$ measures local polarization; by itself it does not distinguish a macroscopic domain from microscopic inhomogeneity. Likewise, vanishing connected fluctuations about an ordered density profile do not imply hyperuniformity of its full density spectrum: the squared Fourier amplitude of the mean profile must be added to $S_c$.

The calculation represents $\rho$ as an MPS over ket--bra supersites and uses time-evolving block decimation (TEBD). Bond dimension adapts from 64 to 512, with singular-value and discarded-weight thresholds $10^{-10}$, trace normalization after each layer, and marginal probabilities below $10^{-14}$ treated as zero. No conserved quantum number is imposed on this density-matrix MPS. The reported maximum dimension is 96 for $b=2$ and reaches 512 transiently for the $b=1$ reference; no run ends at that cap. A validation compares to exact density-matrix evolution at $L=6$, with agreement to $10^{-14}$.

\subsection{Sequential pure-state trajectories}
Figure~\ref{fig:magdata} instead draws each outcome from its own conditional pure state, implementing $P_b^{\rm seq}$ in Eq.~\eqref{eq:ensembles}. Its structure factor retains both the conditional density profiles and their quantum fluctuations.

Here $f_U=0.6$. The unlike-pair block is
\begin{equation}
 D_1\begin{pmatrix}\cos\theta&-i\sin\theta\\-i\sin\theta&\cos\theta\end{pmatrix}D_2,
 \qquad\sin^2\theta=r=0.3,
\end{equation}
with independent uniform diagonal phases in $D_1,D_2$ and independent phases on $|00\rangle,|11\rangle$. After each gate layer, each site is selected independently with probability $f_{\rm m}=0.1$, with readouts processed left to right. Each selected outcome is drawn with probability $p_i(m)^b/\sum_a p_i(a)^b$ from the current pure state, then projected and normalized. All observations are single-site occupation measurements.

For the spectra in Fig.~\ref{fig:magdata}, each $b=0.5,1,2$ and $L=20,24,32,48$ uses 512 trajectories, with common random numbers across exponents. Runs last $T=1.5(L-4)^2$ layers and record observables every ten layers. The plotted quantity is the full structure factor
\begin{equation}
 S(q)=\frac1L\E_{\rm seq}\overline{\langle\psi_t|\delta n_q^\dagger\delta n_q|\psi_t\rangle}.
\end{equation}
Here $\delta n_q=\sum_x(n_x-1/2)e^{-iqx}$. It is not the spectrum of conditional mean densities: it includes their squared Fourier amplitudes and the conditional quantum covariance. At the resolved configurations of the SEP construction, it reduces to Eq.~\eqref{eq:Sqdef}. The dashed reference $1/4$ is the uncorrelated thermodynamic value; the finite-size fixed-number equilibrium value is $L/[4(L-1)]$.

The $L=20$ trajectories use an exact fixed-number state vector. Larger sizes use $U(1)$-conserving MPS TEBD with inversion-free two-site updates and one orthogonality-center sweep for each layer's readouts. Adaptive bond dimension ranges from 64 to 512, with the same $10^{-10}$ truncation thresholds. The reported maximum dimensions are 192, 288, and 512 for $L=24,32,48$. The seed-matched comparison to exact trajectories at $L=8$ agrees to $4\times10^{-15}$. These small-system comparisons validate implementation, but do not replace convergence checks for the largest-size, longest-wavelength modes.

At $b=0.5$, the accessible spectrum decreases approximately as $|q|^{1/2}$ over $0.13\lesssim q\lesssim0.8$. This is evidence for suppression of long-wavelength fluctuations, not an established asymptotic exponent. 

%The data are stored as \texttt{dmtebd\_data/Warp.pkl} for (a,b) and \texttt{dmtebd\_data/p\_Born\_TEBD.pkl} for (c), with plotting script \texttt{analysis/fig\_warped\_orders.py}.

\subsection{Late-time stationarity}
\begin{figure*}[t]
\centering
\includegraphics[width=\textwidth]{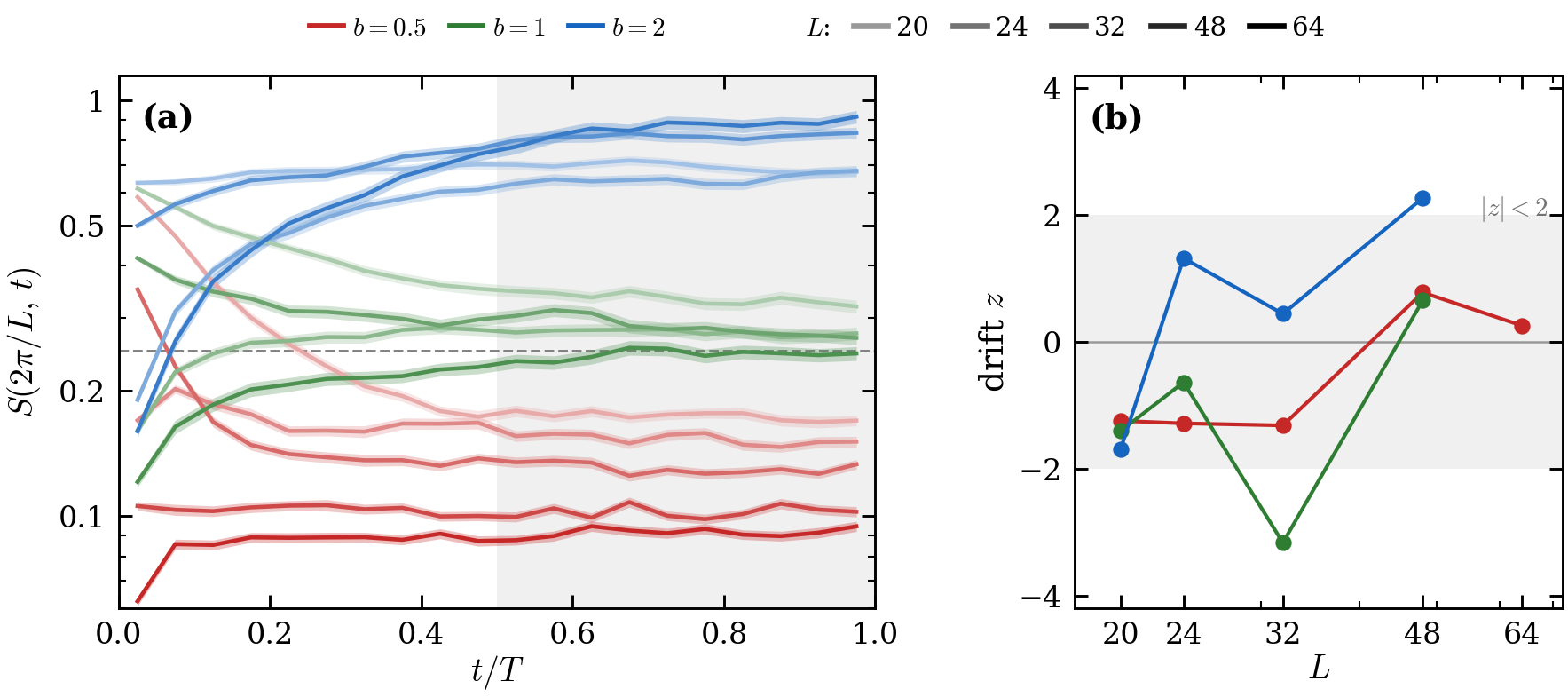}
\caption{Late-time settling in sequential warped-Born histories. (a) The full structure factor $S(2\pi/L,t)$ for $b=0.5,1,2$, averaged over trajectories in 20 equal bins of $t/T$. Shades distinguish sizes; $L=64$ is included only for $b=0.5$. Bands show one standard error, the dashed line is $1/4$, and the grey region marks $t\geq T/2$. (b) The paired drift statistic $z$ compares the last and first thirds of this late-time window. Eleven of thirteen cases fall within $|z|<2$; the two residual drifts are quantified in the text. The $b=2$ runs have twice the duration used for the other exponents at the same size.}
\label{fig:stationarity}
\end{figure*}

We test the settling of the longest-wavelength density fluctuations directly by following $S(q_{\min},t)$, with $q_{\min}=2\pi/L$. Figure~\ref{fig:stationarity} shows late-time plateaus for $b=0.5,1,2$, providing evidence that the clustering and fluctuation suppression persist after the initial relaxation.

The gates, single-site measurements, initial states, and sampling procedure are those of the sequential pure-state protocol above: $f_U=0.6$, $f_{\rm m}=0.1$, and $r=0.3$. The stationarity runs use $T=1.5(L-4)^2$ layers for $b=0.5,1$ and $T=3(L-4)^2$ for $b=2$. Thus the positive-bias test extends the runs used for the spectra in Fig.~\ref{fig:magdata}. Each $(b,L)$ has 512 trajectories for $L=20,24,32,48$; an additional $b=0.5,L=64$ run has 483 trajectories. Observables are recorded every ten layers, with common random numbers across exponents. The state-vector and MPS methods and truncation tolerances are unchanged. The bond-dimension cap of 512 is reached by at most a few trajectories at $L\geq48$ (for example, 2 of 512 at $b=0.5,L=48$).

To quantify drift without treating correlated time samples as independent, split $[T/2,T]$ into three equal intervals. For trajectory $k$, take the difference between its mean structure factor in the last and first intervals,
\begin{align}
 \Delta_k&=\overline{S_k(q_{\min})}_{[5T/6,T]}
          -\overline{S_k(q_{\min})}_{[T/2,2T/3]},\notag\\
 z&=\frac{\E_k\Delta_k}{\operatorname{SEM}_k(\Delta_k)}.
 \label{eq:stationaritydrift}
\end{align}
Here $S_k(q,t)=L^{-1}\langle\psi_{k,t}|\delta n_q^\dagger\delta n_q|\psi_{k,t}\rangle$, and the standard error is taken over independent trajectories. The resulting $z$ values are
\begin{center}
\begin{tabular}{c|rrrrr}
 $b\backslash L$ &20&24&32&48&64\\\hline
 $0.5$&$-1.2$&$-1.3$&$-1.3$&$0.8$&$0.3$\\
 $1$&$-1.4$&$-0.6$&$-3.2$&$0.7$&---\\
 $2$&$-1.7$&$1.3$&$0.4$&$2.3$&---
\end{tabular}
\end{center}
All five $b=0.5$ runs and three of four runs for each of $b=1,2$ have no detectable drift by the plotted $|z|<2$ criterion. The exceptions are a downward residual drift at $b=1,L=32$ ($z=-3.2$) and an upward residual drift at $b=2,L=48$ ($z=2.3$). Together with the plateaus, these tests support late-time settling for all three exponents, with the stated finite-window residuals. They test stationarity of the longest-wavelength structure factor; convergence in bond dimension and the asymptotic small-$q$ scaling are separate checks.

\section{Connections to few-body pathologies}
\label{app:fewbody}
This section relates the many-body density instability to the few-body consequences of modified probability rules. The mechanisms are related, but many operational assumptions differ. Aaronson analyzes the restrictions on linear $2b$-norm-preserving maps and the consequences of relaxing norm preservation~\cite{AaronsonIsland}. We refer readers there for the constructions and derivations.

\textit{Discrimination and information.} Amplifying unequal outcome weights changes how well observations distinguish nonorthogonal alternatives. Our local entropy response expresses the same sharpening of likely outcomes for $b>1$ and weakening of likely outcomes for $b < 1$. Its collective consequence requires conserved transport and a specified history ensemble.

\textit{Grouping outcomes.} The warped rule depends on which alternatives the instrument resolves: merging two Born weights $x,y$ before warping gives $(x+y)^b$, whereas erasing their labels after a fully resolved warped readout gives $x^b+y^b$. For example, in the one-particle state
\begin{equation}
 \ket\psi=\tfrac12\ket{100}+\tfrac12\ket{010}
             +\tfrac1{\sqrt2}\ket{001},
 \label{eq:three_site_obstruction}
\end{equation}
a fully resolved position readout at $b=2$ assigns probabilities $(1/6,1/6,2/3)$, so merging its first two outcomes gives $\Pr(n_3=0)=1/3$. A direct single-site readout of $n_3$ instead gives $1/2$. This comparison concerns different instruments; the dynamics studied here specify single-site readouts and their order.

\textit{Signaling and normalization.} Even that sequential single-site rule violates operational no-signalling if ordinary projection of a shared state and optional remote readouts are allowed. For Eq.~\eqref{eq:three_site_obstruction}, let Alice control site 1 and Bob site 3, with no intervening gate and Alice's readout ordered first. Without Alice's measurement, Bob finds $n_3=1$ with probability $1/2$. If Alice measures and discards her result, Bob's probability becomes
\begin{equation}
 \Pr_b(n_3=1)=\frac{3^b}{1+3^b}\frac{2^b}{1+2^b},
\end{equation}
which is $18/25$ at $b=2$. Thus particle number conservation alone does not prevent the signalling obstruction associated with modified probability rules~\cite{AaronsonIsland}. This is a limitation on interpreting the specified stochastic protocol as a fundamental local measurement theory, not a mechanism needed for its density instability.

\textit{Postselection and computational power.} Enhanced computational power is an explicit conclusion of Aaronson's work: with arbitrary polynomial-size unitary circuits, ancillas, and a final fully resolved $2b$-power readout, every fixed $b\ne1$ permits simulation of $\mathsf{PP}$~\cite{AaronsonPP}. His construction can split an amplitude among $M$ equal orthogonal ancillary branches, changing its summed warped weight by $M^{1-b}$. A coarse single-site projector does not implement this refinement. Applying the computational conclusion to the restricted number-conserving instruments here would therefore require an efficient encoding of those resources; the history bias alone does not establish that encoding.

\section{Replica moments and physical averages}
\label{app:replica}
Here we derive the distinction between the physical Born average and the normalized linear $n$-copy contraction in Eq.~\eqref{eq:replicas}. We then connect that distinction to explicit constructions in the replica literature.

At fixed external protocol the physical $n$-copy state is
\begin{equation}
 R_n^{\rm Born}=\sum_{\mathbf m}P_{\rm B}(\mathbf m)\rho_{\mathbf m}^{\otimes n}.
\end{equation}
A cyclic permutation on a subsystem extracts the Born average of its purity moment $\operatorname{Tr}\rho_A^n$; obtaining an averaged R\'enyi entropy additionally involves the logarithm. Evolving unnormalized copies with a common measurement outcome instead gives
\begin{equation}
 \tilde R_n=\sum_{\mathbf m}P_{\rm B}(\mathbf m)^n\rho_{\mathbf m}^{\otimes n}.
\end{equation}
Normalizing by $\operatorname{Tr}\tilde R_n$ yields the global law at $b=n$, rather than $R_n^{\rm Born}$. External averaging before this normalization gives
\begin{equation}
 \frac{\E_Q\tilde R_n}{\operatorname{Tr}\E_Q\tilde R_n}
 =\int\mathbb P_{b=n}^{\rm ann}(\dd\omega,\mathbf m)\rho_{\mathbf m}^{\otimes n},
\end{equation}
where the integral includes the sum over outcomes. Averaging normalized moments instead retains the original $Q$. The distinction is the same annealed-versus-fixed-protocol distinction as in Appendix~\ref{app:mapping}.

The ratio in Eq.~\eqref{eq:replicas} is a contraction of the linearly evolved $\tilde R_n$ in both numerator and denominator. A cyclic permutation contracts its numerator to a purity moment, while the identity contracts its denominator to $\sum_{\mathbf m}P_{\rm B}^n$. Linearity refers to this replicated unnormalized evolution; taking the final ratio does not turn its weights into $P_{\rm B}$.

For a concrete correspondence, Eq.~(3) of Bao, Choi, and Altman~\cite{BaoChoiAltmanTheory} forms a logarithmic ratio of averaged $n$th moments of the system-plus-record and record density matrices. Resolving their classical record blocks gives the weights $P_{\rm B}(\mathbf m)^n$. Equation~(15) of Leontica and McGinley~\cite{LeonticaMcGinley} displays this outcome weighting explicitly, with normalization after averaging. It also includes a factor $d^{M(n-1)}$ for a schedule with $M$ measurements and local Hilbert dimension $d$. This factor cancels at a fixed schedule, but changes the joint schedule weight when measurement locations are averaged. Their moment ratio therefore shares the conditional outcome tilt while using an additional schedule weighting.

Two further examples make the weighting explicit. Equations~(19)--(20) of Bao, Choi, and Altman's symmetry analysis~\cite{BaoChoiAltmanSymmetry} use $p_{\mathbf m}^2/\sum_{\mathbf m'}p_{\mathbf m'}^2$ in their two-copy observables. Bentsen, Sahu, and Swingle~\cite{BentsenSahuSwingle}, Eq.~(10), evaluate $\langle Z_2\rangle/\langle P^2\rangle$, where $Z_2(V)=P(V)^2\operatorname{Tr}\rho_Q(V)^2$. This weights circuit realizations by $\pi(V)P(V)^2$, with $P(V)$ the success probability of their postselected measurement record and $\pi(V)$ the prescribed circuit distribution. Thus the latter is a joint circuit-weighting example of the same two-copy normalization. Feng et al.~\cite{FengQuantumTrees}, Appendix G, likewise study normalized $P_{\rm B}^n$ record weights on quantum trees and find different spin-sharpening phase boundaries from Born weighting. McGinley, Ho, and Malz~\cite{McGinleyHoMalz2025}, Sec.~IV, also review the ratio of averaged moments defining quasientropy and explain why its fixed-integer values are not Born-averaged entropies; their principal results use a separate, replica-free method. These examples establish the normalization structure; the diffusive spatial density instability studied here is a distinct collective consequence.

Barratt et al.\ explicitly include the normalization and Born factor in their replicated expression, then take the replica number $Q\to1$ to recover physical averages [Eq.~(2) of Ref.~\cite{barratt_field_2022}]. Their field theory explicitly warns that fixed replica number can give qualitatively wrong phases and critical properties; it should not be read as a calculation that omits Born weighting. Fixed integer moments of the normalized common-record construction above retain $P_{\rm B}^n$ instead. Integer-replica ordering consequently need not be a phase of the physical one-copy state. In models admitting the mapping, fixed $n>1$ lies on the positive-bias ($b > 1$) side; taking a large-system limit before approaching the physical limit can therefore select a different ensemble of particle histories within the same fixed-number sector. Replica computations designed to reconstruct Born observables must retain the appropriate normalizers and replica limit~\cite{BaoChoiAltmanTheory,barratt_field_2022,agrawal_entanglement_2022}. A constraint that keeps the common replica density uniform throughout a record removes separated profiles, as motivated in Ref.~\cite{OgunnaikeTower}, but defines a restricted ensemble rather than automatically recovering Born statistics. Uniform mean density and hyperuniform connected fluctuations are distinct diagnostics.
\end{document}